# Nonlinear Optical Susceptibilities and Linear Absorption in Phosphorene Nanoribbons: Ab initio study

Sima Shekarforoush[1], Daryoush Shiri[2*] and Farhad Khoeini*[1]

[1]Department of Physics, Zanjan University, P. O. Box 45195-313, Zanjan, Iran

[2]Department of Physics, Chalmers University of Technology, SE-41296 Göteborg, Sweden

*Corresponding Authors' Email: shiri@chalmers.se, khoeini@znu.ac.ir

**Abstract:** Using Density Functional Theory (DFT) method we compute linear optical absorption spectra and nonlinear optical susceptibilities of hydrogen passivated armchair and zigzag Phosphorous Nanoribbons (aPNR and zPNR) as well as α-phase phosphorous monolayer. We observe that: (a) Crystallographic direction has a strong effect on the band edge absorption which causes optical anisotropy as well as a red shift of absorption spectra by increasing the nanoribbon width. (b) The absorption values are in the order of $10^5$ cm$^{-1}$ which are similar to the experimentally measured values. (c) There is two orders of magnitude enhancement of the 2$^{nd}$ order nonlinear optical susceptibility, $\chi^{(2)}$, in nanoribbons which emanates from breaking the centro-symmetric structure of a monolayer phosphorene by hydrogen surface terminations. (d) Chief among our results is that the 3$^{rd}$ order susceptibility, $\chi^{(3)}$, for phosphorene monolayer and nanoribbons are about $\approx 10^{-13}$ esu ($\approx 10^{-21}$ m$^2$/V$^2$) which are in close agreement with experimentally reported values as well as a recently calculated value based on semi-analytic method. This strongly supports reliability of our method in calculating nonlinear optical susceptibilities of phosphorene and in general other nanostructures. Enhanced 2$^{nd}$ order optical nonlinearity in phosphorene promises better second harmonic and frequency difference (THz) generation for photonics applications.

**Introduction**

Theoretical and experimental research on black phosphorous (BP) or phosphorene have gained significant momentum during recent years due to the interesting electronic, optical, mechanical and thermal properties this new two dimensional materials offer. Unlike graphene which lacks an energy bandgap [1], BP has direct bandgap which is controllable by the number of layers. The medium value of bandgap (0.6eV - 2eV) [2] puts BP between graphene and transition metal dichalcogenides (TMD) [3] and makes it a suitable choice for a high $I_{on}/I_{off}$ ratio field effect transistor (FET). Although the on/off ratios reported so far are in the order of $10^4$ which is less than that of TMD's ($10^6$), still they can provide acceptable performance for analog applications e.g. biosensors [4,5]. The anisotropic band structure of phosphorene around Brillouin Zone (BZ) center results in two different effective mass i.e. different conductance values for electrons moving along zigzag (ZZ) or armchair (AC) directions [6]. Additionally, the anisotropic phonon band structure of phosphorene monolayers leads to anisotropic mechanical properties [7] and phonon-phonon scattering. Hence the thermal conductivity in this two dimensional material is anisotropic [6] although still inferior when it comes to comparison with graphene with thermal conductivity of bout 3000 W/m.K [8]. Vital for the applications in optoelectronic devices, BP shows direct controllable bandgap covering a wide spectrum of photon energies (UV-IR) as well as polarization-dependent absorption i.e. optical anisotropy [9]. This means that a BP layer is transparent to light polarized along the zigzag direction due to symmetry-forbidden optical transitions; on the other hand it is absorptive for the photons polarized along the armchair direction.

Motivated by this, many experimental studies flourished on the linear and nonlinear optical properties of phosphorene in order to bring this material into realm of applications like all-optical switches, limiters, phase modulators, filters, beam-splitters and polarizers among others. To name a few; optical reflectance measurements [10], ultraviolet photoemission spectroscopy (UPS) and X-ray photoemission spectroscopy (XPS), angle-resolved UPS, infrared absorption due to phonons i.e. Raman spectroscopy [11], and measuring phonon dispersion by inelastic neutron scattering [12]. These studies reveal interesting optical properties of BP including dependence of optical absorption intensity on the number of layers, doping, and polarization of photons as well as applied normal electric field [13]. It was observed that application of gate voltage leads to interesting effects like quantum-confined Franz-Keldysh effect and Pauli-blocked Burstein–Moss shift, both of which make α-P BP a useful infrared electro-optic material [14] in sensing, spectroscopic molecular analysis, and free-space optical communications [15,16]. Higher UV absorption in α-P BP compared to graphene makes it more attractive for fabricating thin film solar cells [17].

Experimental studies of nonlinear optical effects revealed interesting physics of carrier dynamics and electron-photon interactions within phosphorene. Although $2^{nd}$ order nonlinear optical susceptibility in bulk phosphorus is small due to its centro-symmetric structure [18], recent studies on odd-layered exfoliated TMDs have estimated strong second harmonic generation (SHG) [19]. This observation and size dependency of $2^{nd}$ order susceptibility suggest that breaking the centro-symmetricity through surface termination and/or size reduction (e.g. making nanoribbons or pellets) will lead to enhanced $2^{nd}$ order nonlinear optical susceptibility, $\chi^{(2)}$. Enhancement of $\chi^{(2)}$ via symmetry breaking was also observed experimentally [20-22] and theoretically [23] for silicon nanowire-based waveguides. Furthermore it was experimentally observed that a large nonlinearity near the band gap can lead to decrease of scattering rates and thermal losses in ultrafast optical modulation [24]. Also the nature of a nonlinear optical process which occurs in phosphorene (e.g. saturable absorption (SA), reverse saturable absorption (RSA) and two photon absorption (TPA)) strongly depends on the input light intensity [25].

Recent experimental studies of $3^{rd}$ order optical nonlinearity suggest that phosphorene [26] and TMDs [27] have higher $3^{rd}$ order susceptibility, $\chi^{(3)}$, compared to graphene at the same wavelength [28]. The measured experimental values [25, 27] of $3^{rd}$ order susceptibility for monolayer phosphorene are in the order of $\sim 10^{-14}$ esu. Also, nonlinear optical properties of black phosphorus quantum dots (BPQDs) were investigated using z-scan technique [29] which shows the $3^{rd}$ order nonlinearity could be enhanced further by embedding the phosphorene dots in liquid. Seeing different experimentally measured values of nonlinear optical susceptibilities and great promise for enhancing the nonlinearity for real-life applications, an atomistic calculation and theoretical understanding of nonlinear optics of phosphorene is necessary.

In this work we used time independent Density Functional theory implemented in SIESTA [30] to calculate linear optical absorption and nonlinear optical susceptibilities [31] of α-phase monolayer phosphorene and phosphorene nanoribbons of zigzag and armchair chirality.

We observed that: *Firstly*, there is a strong anisotropy in the absorption spectra of the phosphorene nanoribbons depending on the chirality as well as polarization of photons. The calculated values are satisfactorily close to the experimental values of $(0.9 - 1) \times 10^5$ cm$^{-1}$. *Secondly*, breaking the centro-symmetry of the nanoribbons due to residual stress as well as hydrogen surface termination results in a 100 fold increase of $\chi^{(2)}$ in nanoribbons as opposed to the monolayer counterpart. *Thirdly*, and importantly, $\chi^{(3)}$ values of phosphorene nanoribbons as well as monolayer phosphorene are in close agreement with the experimentally measured values and an analytically calculated (perturbation-based) value of $10^{-21}$ m$^2$/V$^2$ as reported in [32]. The satisfactory agreement of our numerical results with the experimental measurements validates our adopted method which is less computationally demanding than time dependent DFT (TDDFT) and perturbative methods. Our results promise applications of this new member of 2D materials, phosphorene, in optoelectronic devices.

The rest of this article is organized as follows. The next section (section II) sketches all details of energy minimization, band structure calculation as well as optical calculations using DFT method implemented in SIESTA®. In Section III, the linear absorption spectra as well as nonlinear optical susceptibilities of phosphorene nanoribbons and monolayer phosphorene are presented and discussed, after which a conclusion and speculation on potential applications are presented (Section IV).

**Computational Methods**

**A. Energy Minimization:** The energy minimization, calculation of band structure, absorption spectra and optical dipole polarization were performed using linear scale Time Independent Density Functional Theory (TIDFT) method implemented in SIESTA® [30]. The double-$\zeta$ polarized (DZP) basis set was chosen for expansion of the electron wave function. The exchange-correlation functional is of Generalized Gradient Approximation (GGA) type with Perdew-Burke-Ernzerhof (PBE) pseudo potentials [33] to avoid the bandgap underestimation caused by Local Density Approximation (LDA) functional. The cutoff energy of 360 Ry is used and the Brillouin Zone (BZ) of ribbons and 2D monolayer sheet were sampled with 1×1×9 and 16×2×16 Monkhorst-Pack grids, respectively. The coordinate optimization was performed by Conjugate Gradients (CG) algorithm and the Hellmann–Feynman atomic force tolerance is less than 0.015 eV/Å. The separation between unit cells is more than 20 Å to avoid atomic interactions between periodic replicas of adjacent unit cells.

**B. Electronic Structure:** A monolayer phosphorene has a puckered honeycomb lattice in which the phosphorous (P) atoms form $sp^3$ hybridization with their three nearest neighbors [See Figure 1(a)]. Each primitive cell contains four atoms. The length of the bonds and the angles between them which is obtained after energy minimization are $d_1$= 2.26 Å, $d_2$=2.28 Å, α=95.54 Å and β=102.35 Å. The unit cell length vectors (a =4.4923 Å and b=3.3474 Å) and the value of direct bandgap ($E_g$=0.95 eV) are calculated by DFT. To put these obtained values in perspective with other DFT-based calculations, it is instructive to consider Table. I. Figure 1(b) shows the atomic structure of unit cells for aPNR and zPNR unit cells. Three different widths were created for the calculation of optical absorption. The name of the unit cells are chosen to be the number of dimer P atoms counted on the top view of each unit cell and their corresponding width, length and bandgap values are summarized in Table. II and they are in good agreement with that values reported in reference [34].

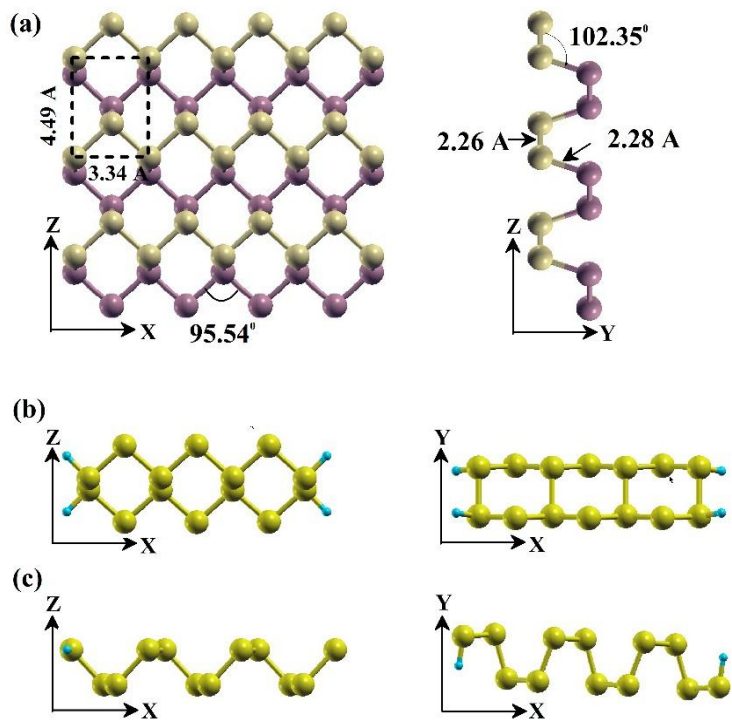

**Figure 1.** (a) Left and right panels depict a section of a 2D monolayer phosphorene. The rectangle encloses a unit cell along with values of bond lengths, angles and unit cell lattice parameters ($d_1$= 2.26 Å, $d_2$=2.28 Å,

α=95.54° and β=102.35°) obtained by SIESTA®. The monolayer lies in XZ plane. (b) The unit cell of an armchair nanoribbon. The ribbon is periodic along Z direction and it is confined along X direction. (c) The unit cell of a zigzag nanoribbon. Left/Right panels show Top/Side views of each unit cell. Yellow and blue atoms are phosphorous and hydrogen, respectively.

The number of diatomic bonds is even and odd for zPNR and aPNR unit cells in order to facilitate the hydrogen termination along the edge of the ribbons. Table II shows that the bandgap value is decreased by increasing the width and approaches to that of 2D monolayer phosphorene as the width grows. This is expected due to quantum confinement or particle in a box model. Figure 2 shows the band structure and density of states of the monolayer, a 12.74 Å wide zigzag nanoribbon (6-zPNR) and a 10.04 Å wide armchair nanoribbon (7-aPNR), respectively.

| Method | a (Å) | b (Å) | $E_g$ (eV) | References |
|---|---|---|---|---|
| PBE (SIESTA®) | 4.4923 | 3.3474 | 0.95 | This Work |
| PBE | 4.627 | 3.298 | 0.92 | Peng et al. [36] |
| OptB88-vdW | 4.506 | 3.303 | 0.76 | Sa et al. [37] |
| HSE062@PBE | 4.627 | 3.298 | 1.54 | Sa et al. [37] |
| HSE062@optB88-vdW | 4.58 | 3.32 | 1.51 | Qiao et al. [9] |
| LDA_Mbj@optB88-vdW | 4.58 | 3.32 | 1.41 | Qiao et al. [9] |
| ModifiedHSE06@PBE | 4.62 | 3.35 | 1.0 | Liu et al. [38] |
| $G_0W_0$@PBE | 4.627 | 3.298 | 2.08 | Sa et al. [37] |
| $G_0W_0$@PBE | 4.52 | 3.31 | 1.94 | Liang et al. [39] |
| $G_0W_0$@PBE_vdW | - | - | 2.0 | Tran et al. [2] |
| BSE($G_0W_0$/ $G_1W_1$) | - | - | 1.2/1.4 | Tran et al. [2] |
| GW | - | - | 1.60 | Rudenko et al [34] |

**Table I.** Comparison of lattice parameters of monolayer phosphorene unit cell and the direct bandgap ($E_g$) values obtained from our DFT method with those based on using other exchange functionals. Table. I is extracted and summarized from reference [35] to facilitate the comparison of our results with different groups.

| System | Width (Å) | Length (Å) | $E_g$ (eV) |
|---|---|---|---|
| 4-zPNR | 8.41 | | 2.11 |
| 6-zPNR | 12.74 | 3.32 | 1.68 |
| 8-zPNR | 17.10 | | 1.43 |
| 5-aPNR | 6.71 | | 1.31 |
| 7-aPNR | 10.04 | 4.27 | 1.12 |
| 9-aPNR | 13.38 | | 1.03 |

**Table II.** Width and length of the zigzag and armchair PNRs as well as the value of direct bandgap ($E_g$) obtained after energy minimization in SIESTA®.

It is evident that monolayer phosphorene has a direct bandgap, i.e., both maximum and minimum of valence and conduction bands are at the same value of wave vector which is $\Gamma$ or Brillouin Zone (BZ) center. The direct bandgap value of 0.95 eV and its controllability with electric field [40, 41] offer possibility of using phosphorene monolayers as electronic transistors with large $I_{on}/I_{off}$ ratio. A high anisotropy in effective mass results from the sharp contrast of sub band curvatures around $\Gamma$ point (e.g. $\Gamma X$ and $\Gamma Y$). Due to the folding of off-center states of 2D phosphorene to the $\Gamma$ point in 1D BZ, the nanoribbons have direct bandgap and an effective mass for conduction band which depends on the chirality. For example cutting the monolayer along $\Gamma X$ to make a zigzag nanoribbon folds the states to $\Gamma$ and make the band structure look like Figure 2(b). The same is true for Figure 2(c) in which cutting an armchair nanoribbon from a monolayer means cutting the bands along $\Gamma Y$ direction and folding them. This leads to a lower effective mass. Anisotropy in effective mass adds flexibility in tuning the transport properties of electronic devices based on phosphorene. The van Hove like singularities on the band edge are due to low curvature (large effective mass) valence and conduction sub bands.

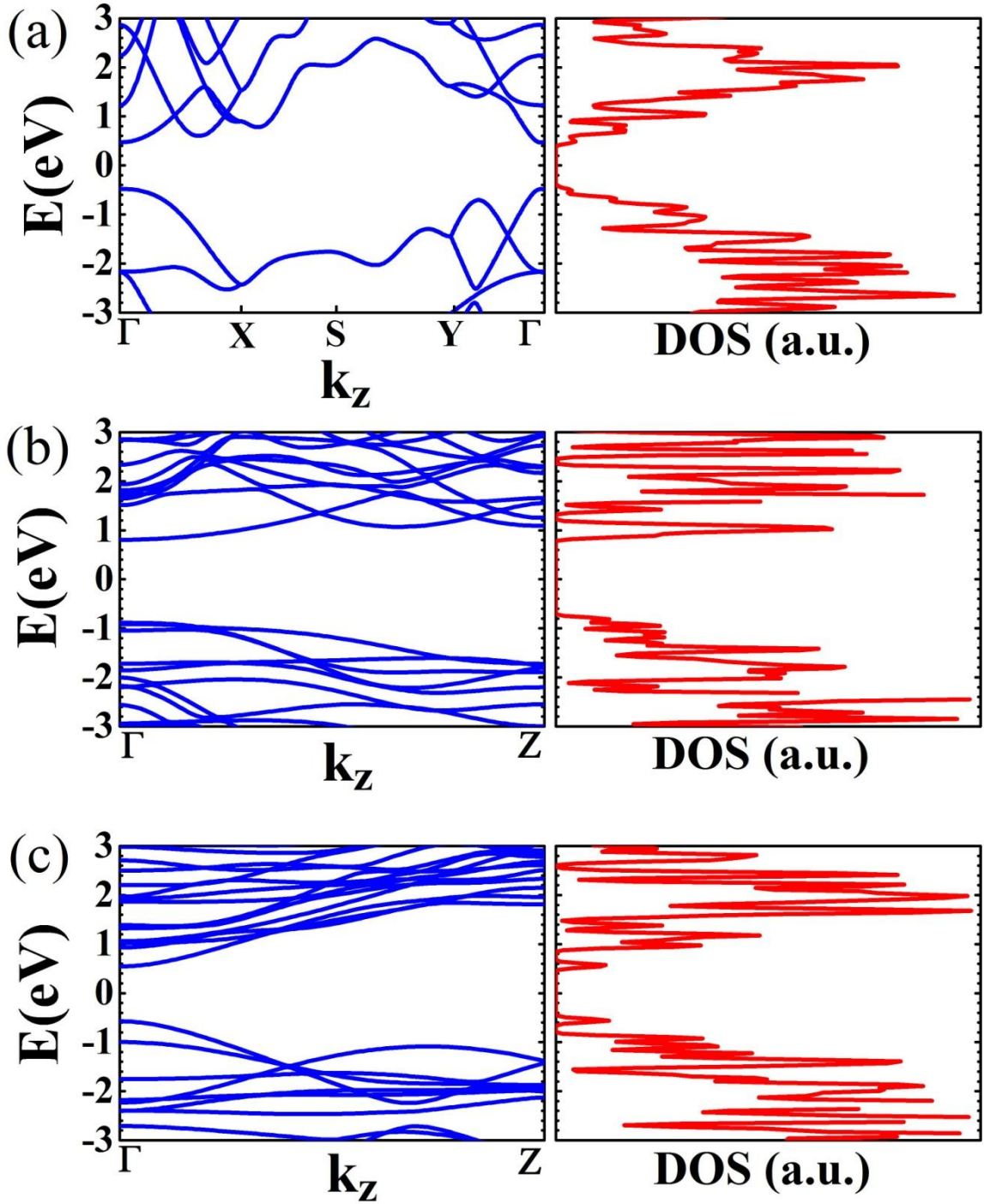

**Figure 2.** Band structures and density of states for (a) monolayer phosphorene, (b) hydrogen terminated 6-zPNR and (c) 7-aPNR. The bandgap values are 0.95 eV, 1.68 eV and 1.12 eV, respectively.

**C. Linear Absorption:** The optical absorption spectra of phosphorene monolayer and nanoribbons are calculated using optical functions implemented in SIESTA® [30]. The real part ($\varepsilon_1$) and imaginary part ($\varepsilon_2$) of dielectric function, $\varepsilon(\omega)$, are necessary to calculate the extinction ratio, $\kappa(\omega)$:

$$\kappa(\omega) = \sqrt{(\sqrt{\varepsilon_1^2 + \varepsilon_2^2} - \varepsilon_1)/2}, \quad (1)$$

from which the absorption, α(ω), is found by:

$$\alpha(\omega) = \frac{4\pi\kappa(\omega)}{\lambda_0} = \frac{4\pi\kappa(\omega)}{c/f} = \frac{2\omega\kappa(\omega)}{c} \quad (2)$$

Here $\lambda$, $c$ and $f$ are wavelength, frequency and speed of light in vacuum, respectively. Corresponding to each electric field polarization (x, y, z), three different spectra for absorption are obtained. The broadening energy and scissor operators are 40 meV and 1 eV, respectively.

**D. Nonlinear Optical Susceptibility:**

Calculation of nonlinear susceptibilities with DFT is based on the calculation of polarization, **P**, in response to the applied electric field, **E**. Since this method is based on the dipole perturbation induced by electric field which is

$$H_{pert} = -\vec{E}.\vec{r} \quad (3)$$

Therefore it is not applicable to periodic systems e.g. nanoribbons and 2D monolayers. However if the electric field is parallel to the smallest non-periodic dimensions of the system, some elements of susceptibility tensor can be extracted. DFT-based methods are computationally more advantageous than perturbation-based methods since there is no need to perform integration over many intermediate states which renders the perturbative methods very time consuming. On the other hand it is well known that DFT method is reliable only at giving the correct ground state of a nano-system although this problem is circumvented by using time-dependent DFT method [42].

**Time-dependent DFT (TDDFT):** Although we have not used this method in our calculations, it is instructive to briefly mention its working principle. In TDDFT, the electric field is assumed to be a time dependent function e.g. a sharp pulse. After the calculation of the ground state at $t_0$=0, the next sample of electric field, **E(t₀+Δt)**, is inserted in perturbation as $H_{pert}(\Delta t) = -\vec{E}(\Delta t).\vec{r}$. Thereafter the electron density is updated self-consistently (similar to the time-independent case) and time-dependent polarization, **P(t)**, is found by:

$$P(t) = \int d^3\vec{r}\rho(\vec{r},t)\vec{r} \quad (4)$$

The polarization, **P(t)**, includes all orders of nonlinearity in response to E(t). The susceptibilities can then be extracted from the Fourier transform of **P(t)**, i.e., **P(ω)** [42].

**Time-independent DFT (TIDFT):** In this method the polarization (**P**) is calculated at each DFT step as a function of static electric field (**E**) using SIESTA®[30, 31]. In general P has three components each of which is a function of electric field components along x, y and z directions. Generally the tensor relation between P and E is written as follows [43]:

$$\vec{P} = P_0 + \epsilon_0 \chi^{(1)} \vec{E} + \epsilon_0 \chi^{(2)} \vec{E}.\vec{E} + \epsilon_0 \chi^{(3)} \vec{E}.\vec{E}.\vec{E} \quad (5)$$

where $P_0$ is static polarization (under zero electric field), $\chi^{(1)}$, $\chi^{(2)}$ and $\chi^{(3)}$ are second, third and fourth rank tensors, respectively i.e. each of which has $3^2=9$, $3^3=27$ and $3^4=81$ components. For phosphorene monolayer the electric field cannot have nonzero component along x and z direction. Hence the electric field is normal to the XZ plane and its component in y direction ($E_y$) can be changed to calculate $P_x$, $P_y$ and $P_z$. As nano ribbons are periodic along z direction, only Ey and Ex can be nonzero. In this case Px, Py and Pz are plotted by changing Ex and Ey and assuming that $E_z$=0. Therefore in the first DFT step the ground state of the energy minimized unit cell is calculated. It is further assumed that the frequency of electromagnetic field (light) and its second and third harmonics are smaller than the bandgap ($E_g$), so there in no electronic excitation. On the other hand these frequencies must be higher than the vibrational frequencies of the nucleus [31]. As an example we expand equation (5) to derive the y component of polarization for unit cell of a zPNR [See Figure 1(b)]:

$$P_y = P_{0y} + \epsilon_0 \left( \chi^{(1)}_{yx} E_x + \chi^{(1)}_{yy} E_y + \chi^{(1)}_{yz} E_z \right) + \epsilon_0 \left( \chi^{(2)}_{yxx} E_x E_x + \chi^{(2)}_{yxy} E_x E_y + \chi^{(2)}_{yxz} E_x E_z + \right.$$
$$\chi^{(2)}_{yyx} E_y E_x + \chi^{(2)}_{yyy} E_y E_y + \chi^{(2)}_{yyz} E_y E_z + \chi^{(2)}_{yzx} E_z E_x + \chi^{(2)}_{yzy} E_z E_y + \chi^{(2)}_{yzz} E_z E_z \right) +$$
$$\epsilon_0 \{ \chi^{(3)}_{yxxx} E_x E_x E_x + \chi^{(3)}_{yxxy} E_x E_x E_y + \chi^{(3)}_{yxxz} E_x E_x E_z + \chi^{(3)}_{yxyx} E_x E_y E_x + \chi^{(3)}_{yxyy} E_x E_y E_y +$$
$$\chi^{(3)}_{yxyz} E_x E_y E_z + \chi^{(3)}_{yxzx} E_x E_z E_x + \chi^{(3)}_{yxzy} E_x E_z E_y + \chi^{(3)}_{yxzz} E_x E_z E_z + \chi^{(3)}_{yyxx} E_y E_x E_x +$$
$$\chi^{(3)}_{yyxy} E_y E_x E_y + \chi^{(3)}_{yyxz} E_y E_x E_z + \chi^{(3)}_{yyyx} E_y E_y E_x + \chi^{(3)}_{yyyy} E_y E_y E_y + \chi^{(3)}_{yyyz} E_y E_y E_z +$$
$$\chi^{(3)}_{yyzx} E_y E_z E_x + \chi^{(3)}_{yyzy} E_y E_z E_y + \chi^{(3)}_{yyzz} E_y E_z E_z + \chi^{(3)}_{yzxx} E_z E_x E_x + \chi^{(3)}_{yzxy} E_z E_x E_y +$$
$$\chi^{(3)}_{yzxz} E_z E_x E_z + \chi^{(3)}_{yzyx} E_z E_y E_x + \chi^{(3)}_{yzyy} E_z E_y E_y + \chi^{(3)}_{yzyz} E_z E_y E_z + \chi^{(3)}_{yzzx} E_z E_z E_x +$$
$$\chi^{(3)}_{yzzy} E_z E_z E_y + \chi^{(3)}_{yzzz} E_z E_z E_z \} \quad (6)$$

Based on the above discussions, as then nanoribbon is periodic along z direction, $E_z$ must be zero i.e. $E_z$=0. Now $P_y$ is a high order polynomial of $E_x$ and $E_y$ which is now:

$$P_y(E_x, E_y) = P_{0y} + \epsilon_0\left(\chi^{(1)}_{yx}E_x + \chi^{(1)}_{yy}E_y\right) + \epsilon_0\left(\chi^{(2)}_{yxx}E_xE_x + \chi^{(2)}_{yxy}E_xE_y + \chi^{(2)}_{yyx}E_yE_x + \chi^{(2)}_{yyy}E_yE_y\right) + \epsilon_0\{\chi^{(3)}_{yxxx}E_xE_xE_x + \chi^{(3)}_{yxxy}E_xE_xE_y + \chi^{(3)}_{yxyx}E_xE_yE_x + \chi^{(3)}_{yxyy}E_xE_yE_y + \chi^{(3)}_{yyxx}E_yE_xE_x + \chi^{(3)}_{yyxy}E_yE_xE_y + \chi^{(3)}_{yyyx}E_yE_yE_x + \chi^{(3)}_{yyyy}E_yE_yE_y\} \quad (7)$$

As can be seen from equation (7) after changing Ex and Ey and plotting P_y, the susceptibility tensor components are extracted using partial derivatives of P_y with respect to E_x and E_y. For example:

$$\chi^{(2)}_{yyx} = \frac{\partial^2 P_y(E_x, E_y)}{\partial y \partial x} \quad , \chi^{(3)}_{yxxy} = \frac{\partial^3 P_y(E_x, E_y)}{\partial^2 x \partial y} \quad (8)$$

However to further simplify and speed up the calculations we only cut the multi-dimensional surface of P_y(E_x, E_y) along its axis by once assuming E_x=0 and E_y≠0 which yields:

$$P_y(0, E_y) = P_{0y} + \epsilon_0\left(\chi^{(1)}_{yy}E_y\right) + \epsilon_0\left(\chi^{(2)}_{yyy}E_yE_y\right) + \epsilon_0\{\chi^{(3)}_{yyyy}E_yE_yE_y\} \quad (9)$$

And in another run we assume E_y=0 and E_x≠0 which returns:

$$P_y(E_x, 0) = P_{0y} + \epsilon_0\left(\chi^{(1)}_{yx}E_x\right) + \epsilon_0\left(\chi^{(2)}_{yxx}E_xE_x\right) + \epsilon_0\{\chi^{(3)}_{yxxx}E_xE_xE_x\} \quad (10)$$

The other components of **P** i.e. P_x and P_z are also found similarly. Fitting equations (9) and (10) to polynomials in MATLAB ® is then a straightforward task from which the nonzero tensor components of χ$^{(2)}$ and χ$^{(3)}$ are extracted.

**RESULTS and DISCUSSIONS**

**Linear Optical Absorption:** Strong optical anisotropy in monolayer phosphorene is observed by closely examining the optical absorption spectra of three different photon polarizations as shown in Figure 3 (a). Figure 3(b) and (c) show imaginary and real parts of dielectric function of the monolayer for three different photon polarizations. For a monolayer subject to x-polarized photon which is in parallel with zigzag direction, the absorption vanishes as it is symmetry forbidden. However it starts after 3.5 eV. This observation is in agreement with reference [9] suggesting that valence and conduction bands along zigzag direction are of the same parity. This results in odd symmetry for the matrix element hence zero value for the integral of type:

$\langle \Psi_c(r)|r|\Psi_v(r)\rangle = \int \Psi_c^*(r) r \Psi_v(r) d^3r \quad (11)$

where $\Psi_c$ and $\Psi_v$ are conduction and valence band wave functions, and **r** is the position operator. Now if the photon polarization is along y direction, the absorption is negligible again as the thickness of the sheet (which is along y direction) is extremely thin (2.14 Å) and there is insignificant interaction between the atoms and electric field unless at high energies e.g. above 7eV in Figure 3(a). However, there is a strong symmetry allowed peak on the band edge for z-polarized photon (red spectrum in Figure 3). This means that armchair and zigzag direction show strong optical anisotropy emanating from symmetry of wave functions along these directions. When photon polarization is rotated by 45 degrees as shown in Figure 3(d), then as it is surmised the band edge absorption drops by 50 % because 50 % of the polarization is along the symmetry forbidden direction (zigzag) with almost zero contribution [See Inset of Figure 3(a)].

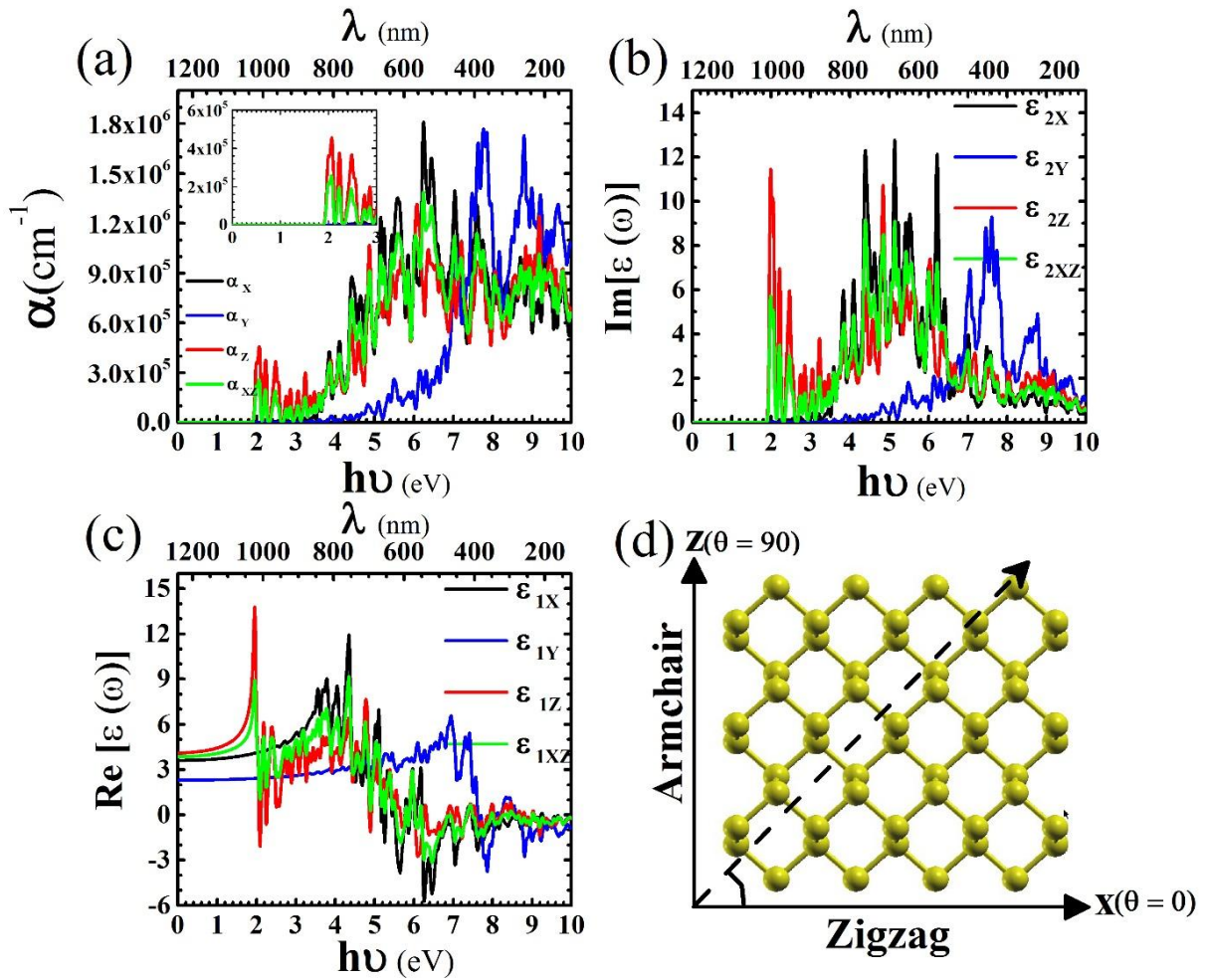

**Figure 3.** (a) The optical absorption spectra, (b) imaginary part and (c) real part of the dielectric function for a monolayer phosphorene. (d) Top view of the relaxed monolayer structure. Inset shows the band edge absorption for three different photon polarizations. The absorption is higher for Z-polarized case, i.e., along armchair edge. For diagonal polarization (XZ), the absorption drops by 50 % since half of the contribution comes from x-polarization which is zero.

For phosphorene nanoribbons the quantum confinement effect is evident by examining the total absorption spectra of armchair and zigzag nanoribbons in Figure 4(a) and 4(b), respectively. In both cases, increasing the width causes red shift of the band edge absorption spectrum since the bandgap value decreases according to Table II. Total absorption spectrum is summation of the three spectra corresponding to each photon polarization. Figure 4(c) and (d) show the absorption spectra of x, y and z-polarized photon for 7-aPNR and 6-zPNR, respectively. The absorption values of all structures around the band edge is agreeably close to the experimentally reported values of $(0.9-1) \times 10^5$ cm$^{-1}$ for phosphorene monolayers and quantum dots [44].

It is noteworthy that no absorption peak is observed for y-polarized cases in both nanoribbons because as mentioned before the thickness is small (2.14 Å) in this direction. For z-polarized photon, 7-aPNR shows strong peaks as this polarization is parallel to armchair direction and the same physics as explained for optical anisotropy in monolayer phosphorene applies here. On the other hand, 6-zPNR has strong absorption peaks for x-polarized photons as x is parallel to the armchair direction for this nanoribbon. The calculated optical anisotropy agrees well with the experimental observations of [45-47]. The transparency of nanoribbon along one direction (zigzag) and their strong optical absorption along other chirality (armchair) promise application of phosphorene in polarization sensitive photo detectors. Adjustability of the band edge absorption with size could be also exploited to broaden the absorption spectrum of photo detectors and solar cells using a matrix or array of phosphorene nanoribbons with different widths.

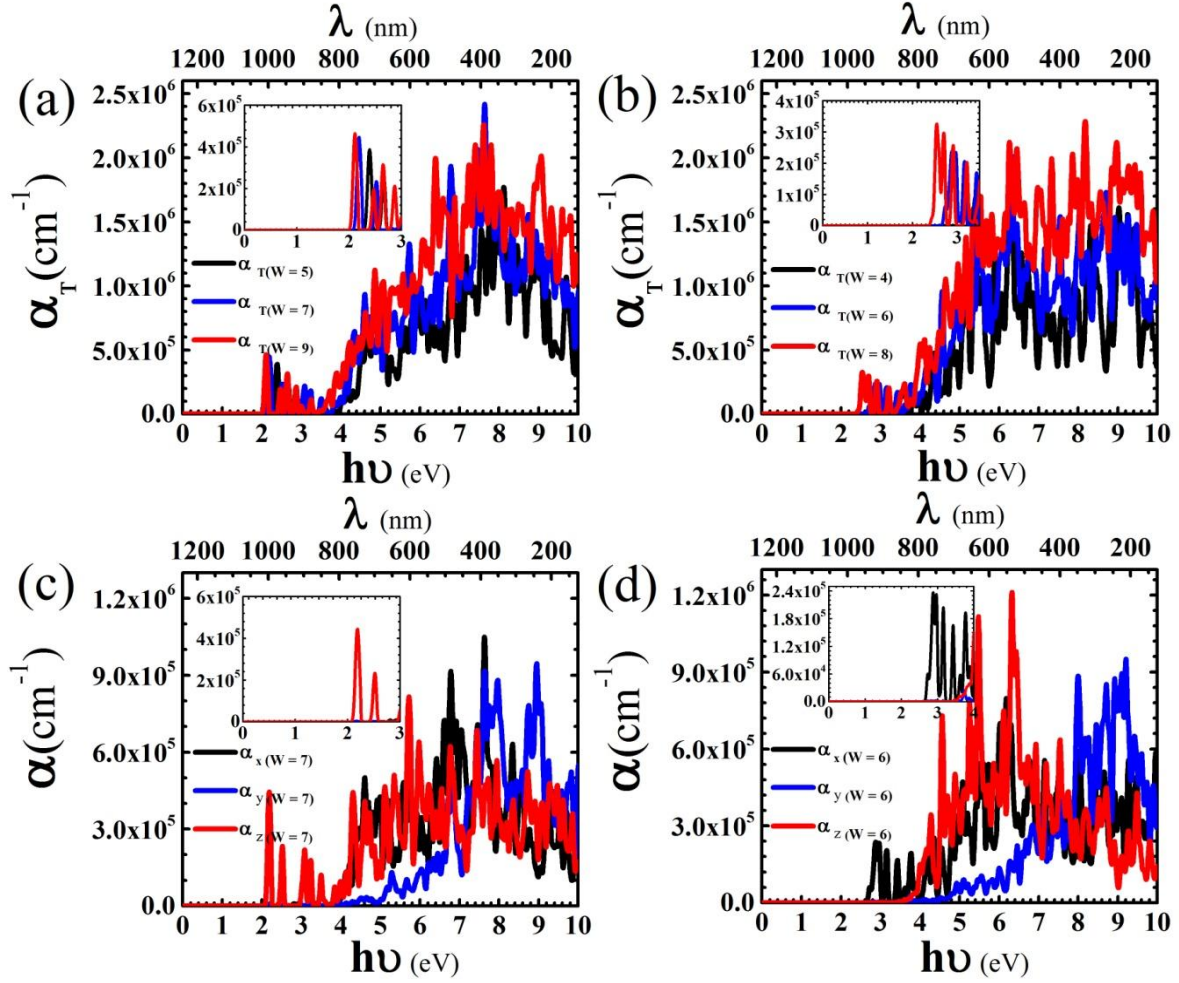

**Figure 4.** Total optical absorption spectra for (a) three different widths of hydrogen passivated aPNRs and (b) zPNRs. The absorption spectra of 7-aPNR (c) and 6-zPNR (d) for three different photon polarizations (x, y & z).

**Nonlinear Optical Susceptibilities:** As mentioned before, in case of phosphorene monolayer, we have two periodic directions (x and z) and the electric field is only applicable along y-direction. Therefore only three tensor components are extractable. The non-zero tensor components of $\chi^{(2)}$ and $\chi^{(3)}$ are summarized in Table III and IV, respectively. As the monolayer phosphorene is of centro-symmetric type [18], the 2$^{nd}$ order nonlinear susceptibility is very small, e.g., $\chi^{(2)}_{yyy}$ = 0.053 pm/V. In contrast to this, by confining the monolayer into nanoribbons and breaking the symmetry due to surface reconstruction, residual stress, as well as sudden surface termination with hydrogen atoms, the 2$^{nd}$ order susceptibility is enhanced by one and two orders of magnitude. For example for 6-zPNR, $\chi^{(2)}_{yyy}$ and $\chi^{(2)}_{xyy}$ are -0.184 pm/V and -1.72 pm/V, respectively. For 7-aPNR, a tenfold enhancement is observed for *xxx* component which is 0.125 pm/V as opposed to 0.053 pm/V for the monolayer phosphorene. This observation suggests that by breaking the centro-symmetric structure of a phosphorene monolayer, the 2$^{nd}$ order nonlinear effects can be enhanced up to 100 times. Converting the metric units of 2$^{nd}$ order and 3$^{rd}$ order susceptibilities to electro-static unit (esu) is straightforward using conversion factors of (1 esu = 4.192×10$^{-4}$ m/V) and (1esu = 1.398×10$^{-8}$ m$^2$/V$^2$), respectively.

|  | 6-zPNR | | 7-aPNR | |
| --- | --- | --- | --- | --- |
|  | pm/V | esu | pm/V | esu |
| $\chi^{(2)}_{xxx}$ | -1.72 | -0.41×10$^{-8}$ | 1.25×10$^{-1}$ | 2.98×10$^{-10}$ |
| $\chi^{(2)}_{yxx}$ | 8.65×10$^{-2}$ | 2.06×10$^{-10}$ | 9.74×10$^{-6}$ | 2.32×10$^{-14}$ |
| $\chi^{(2)}_{zxx}$ | 5.44×10$^{-4}$ | 1.30×10$^{-12}$ | -2.68×10$^{-4}$ | -6.39×10$^{-13}$ |
| $\chi^{(2)}_{yyy}$ | -1.84×10$^{-1}$ | -0.44×10$^{-9}$ | -1.99×10$^{-3}$ | -4.75×10$^{-12}$ |
| $\chi^{(2)}_{xyy}$ | -4.32×10$^{-1}$ | -1.03×10$^{-9}$ | 1.10×10$^{-3}$ | 2.62×10$^{-12}$ |
| $\chi^{(2)}_{zyy}$ | -1.66×10$^{-3}$ | -0.40×10$^{-11}$ | 8.22×10$^{-4}$ | 1.96×10$^{-12}$ |
| $\chi^{(2)}_{yyy}$ | **For monolayer phosphorene where both E$_x$ and E$_z$ are zero** **-5.3×10$^{-2}$ pm/V = -1.27×10$^{-14}$ esu** | | | |

**Table III**. Non-zero 2$^{nd}$ order nonlinear susceptibilities of nanoribbons (6-zPNR & 7-aPNR) and a phosphorene monolayer.

The calculated 3$^{rd}$ order susceptibilities reveal another interesting feature in phosphorene nanoribbons. As it is observed the monolayer phosphorene offers a value of $\chi^{(3)}_{yyyy}$ = 3.52×10$^{-22}$ m$^2$/V$^2$ (or 2.52×10$^{-14}$ esu). It is noteworthy that this value is in close agreement with the experimentally observed measurements of the 3$^{rd}$ order susceptibility reported by Z-scan method and values extracted from the measurement of two photon absorption (TPA). The experimental values vary from -0.5×10$^{-14}$ esu to 15×10$^{-14}$ esu, depending on the wavelength used in measuring the TPA coefficient. As it is observed in Table. IV, the diagonal elements of χ$^{(3)}$ tensor i.e. *xxxx* and *yyyy* are enhanced to the order of 10$^{-21}$ m$^2$/V$^2$ which closely agrees with the values calculated by [32]. It is also noteworthy that diagonal 3$^{rd}$ order susceptibility values (i.e. *xxxx* and *yyyy*) for 6-zPNR are one order of magnitude higher than those of monolayer phosphorene. For example $\chi^{(3)}_{yyyy}$ is 4.09x10$^{-21}$ m$^2$/V$^2$ (or 2.93×10$^{-13}$ esu) as opposed to 2.52x10$^{-14}$ esu for a monolayer.

|  | 6-zPNR | | 7-aPNR | |
|---|---|---|---|---|
|  | m$^2$/V$^2$ | esu | m$^2$/V$^2$ | Esu |
| $\chi^{(3)}_{xxxx}$ | 4.09×10$^{-21}$ | 2.93×10$^{-13}$ | 7.37×10$^{-24}$ | 5.27×10$^{-16}$ |
| $\chi^{(3)}_{yxxx}$ | -3.75×10$^{-22}$ | -2.68×10$^{-14}$ | -5.59×10$^{-25}$ | -3.40×10$^{-17}$ |
| $\chi^{(3)}_{zxxx}$ | -2.25×10$^{-24}$ | 1.61×10$^{-16}$ | -3.15×10$^{-25}$ | -2.25×10$^{-17}$ |
| $\chi^{(3)}_{yyyy}$ | 1.46×10$^{-21}$ | 1.04×10$^{-13}$ | 1.35×10$^{-22}$ | 1.89×10$^{-14}$ |
| $\chi^{(3)}_{xyyy}$ | -1.20×10$^{-23}$ | -0.86×10$^{-15}$ | 3.91×10$^{-25}$ | 2.80×10$^{-17}$ |
| $\chi^{(3)}_{zyyy}$ | -6.69×10$^{-26}$ | -4.79×10$^{-18}$ | 2.44×10$^{-24}$ | 1.74×10$^{-16}$ |
| $\chi^{(3)}_{yyyy}$ | For monolayer phosphorene where both E$_x$ and E$_z$ are zero 3.53×10$^{-22}$ m$^2$/V$^2$ = 2.52×10$^{-14}$ esu | | | |

**Table IV.** Non-zero 3$^{rd}$ order nonlinear susceptibilities of nanoribbons (6-zPNR & 7-aPNR) and a phosphorene monolayer.

With respect to silicon, the 3$^{rd}$ order susceptibility of phosphorene is much smaller than the same quantity for bulk silicon and silicon nanowires, which is about (0.1-1)×10$^{-18}$ m$^2$/V$^2$. Experimental measurements of $\chi^{(3)}$ for graphene, MoS$_2$, MoSe$_2$ and MoTe$_2$ using Z-scan technique [48] reveals that this quantity is in the order of 10$^{-14}$ - 10$^{-15}$ esu for these materials. Further enhancement of $\chi^{(3)}$ in phosphorene could be envisaged using higher number of layers, dispensing phosphorene nano pellets or (quantum dots ) in liquid solution or a matrix of nanoribbons with mixed chirality.

**CONCLUSIONS**

We investigated the linear and nonlinear optical properties of monolayer and hydrogen terminated nanoribbons of phosphorene using a time independent DFT method implemented in SIESTA®. Vital for the application of phosphorene in photodetectors we observed that the band edge absorption of phosphorene is very anisotropic and depends on the incident photon polarization. Additionally, the band edge absorption shows a red shift by increasing the nanoribbon width in concordance with quantum confinement effect on the bandgap value. The absorption values are in the range of 1x10$^5$ cm$^{-1}$ which are very close to the experimentally reported values for phosphorene nano pellets and monolayers. These results suggest application of phosphorene in spectrum widening of photocells and polarization sensitive photodetectors.

By recording the electric dipole polarization in response to the applied electric field using DFT-based method, we extracted the 2$^{nd}$ and 3$^{rd}$ order nonlinear susceptibilities. The importance of the results is twofold. ***Firstly*** we show our adopted computational method return $\chi^{(3)}$ values in close agreement with those reported in experiments using various methods e.g. Z-scan and two photon absorption (TPA) measurement as well as theoretical calculations using a semi-analytic method. Although all components of susceptibility tensors are not extractable in our method however a fast and reliable approximation of these values is offered by this method without resorting to complex and computationally demanding time dependent DFT methods.

***Secondly*** the enhanced values of $\chi^{(2)}$ in nanoribbons suggest that the centro-symmetricity of monolayer phosphorene can be broken using surface termination. This promises new applications of phosphorene nano-structures in second harmonic generation, THz sources, optical switches, and modulators.

**Acknowledgements**

Daryoush Shiri acknowledges Dr. Andrei Buin, University of Toronto, Canada, for his helpful suggestions at the beginning of this work.

**References**

[1] K. Kim, J.-Y. Choi, T. Kim, S.-H. Cho, and H.-J. Chung, Nature **479**, 338 (2011).

[2] V. Tran, R. Soklaski, Y. Liang, and L. Yang, Phys. Rev. B **89**, 235319 (2014).

[3] K. F. Mak and J. Shan, Nature Photonics **10**, 216 (2016).

[4] Q. H. Wang, K. Kalantar-Zadeh, A. Kis, J. N. Coleman, and M. S. Strano, Nature Nanotechnology **7**, 699 (2012).

[5] L. Li, Y. Yu, G. J. Ye, Q. Ge, X. Ou, H. Wu, D. Feng, X. H. Chen, and Y. Zhang, Nature Nanotechnology **9**, 372 (2014).

[6] S. Lee, F. Yang, J. Suh, S. Yang, Y. Lee, G. Li, H. Sung Choe, A. Suslu, Y. Chen, C. Ko, J. Park, K. Liu, J. Li, K. Hippalgaonkar, J. J. Urban, S. Tongay, and J. Wu, Nature Communications **6**, 8573 (2015).

[7] H. Chen, P. Huang, D. Guo, and G. Xie, The Journal of Physical Chemistry C **120**, 29491 (2016).

[8] E. Pop, V. Varshney, and A. K. Roy, MRS Bulletin **37**, 12731281 (2012).

[9] J. Qiao, X. Kong, Z.-X. Hu, F. Yang, and W. Ji, Nature Communications **5**, 4475 (2014).

[10] T. Takahashi, K. Shirotani, S. Suzuki, and T. Sagawa, Solid State Communications **45**, 945 (1983).

[11] M. Ikezawa, Y. Kondo, and I. Shirotani, Journal of the Physical Society of Japan **52**, 1518 (1983).

[12] Y. Fujii, Y. Akahama, S. Endo, S. Narita, Y. Yamada, and G. Shirane, Solid State Communications **44**, 579 (1982).

[13] T. Low, A. S. Rodin, A. Carvalho, Y. Jiang, H. Wang, F. Xia, and A. H. Castro Neto, Phys. Rev. B **90**, 075434 (2014).

[14] C. Lin, R. Grassi, T. Low, and A. S. Helmy, Nano Letters **16**, 1683 (2016).

[15] R. Soref, Nature Photonnics **4**, 495 (2010).

[16] R. Stanley, Nature Photonics **6**, 409 (2012).

[17] J. Dai and X. Zeng, Journal of Physical Chemistry Letters **5**, 1289 (2014).

[18] J. Ribeiro-Soares, R. M. Almeida, L. G. Cancado, M. S. Dresselhaus, and A. Jorio, Phys. Rev. B **91**, 205421 (2015).

[19] L. M. Malard, T. V. Alencar, A. P. M. Barboza, K. F. Mak, and A. M. de Paula, Phys. Rev. B **87**, 201401 (2013).

[20] R. S. Jacobsen, K. N. Andersen, P. I. Borel, J. Fage- Pedersen, L. H. Frandsen, O. Hansen, M. Kristensen, A. V. Lavrinenko, G. Moulin, H. Ou, C. Peucheret, B. Zsigri, and A. Bjarklev, Nature **441**, 199 (2006).

[21] M. Cazzanelli, F. Bianco, E. Borga, G. Pucker, M. Ghulinyan, E. Degoli, E. Luppi, V. Veniard, S. Ossicini, D. Modotto, S. Wabnitz, R. Pierobon, and L. Pavesi, Nature Materials 11, 148 (2012).

[22] B. Chmielak, M. Waldow, C. Matheisen, C. Ripperda, J. Bolten, T. Wahlbrink, M. Nagel, F. Merget, and H. Kurz, Opt. Express **19**, 17212 (2011).

[23] D. Shiri, https://arxiv.org/abs/1707.08324 (2017).

[24] S. Yu, X. Wu, Y. Wang, X. Guo, and L. Tong, Advanced Materials **29**, 1606128 (2017).

[25] R. Chen, Y. Tang, X. Zheng, and T. Jiang, Appl. Opt. **55**, 10307 (2016).

[26] G. Wang, S. Zhang, X. Zhang, L. Zhang, Y. Cheng, D. Fox, H. Zhang, J. N. Coleman, W. J. Blau, and

J. Wang, Photon. Res. **3**, A51 (2015).

[27] J. Zhang, X. Yu, W. Han, B. Lv, X. Li, S. Xiao, Y. Gao, and J. He, Opt. Lett. **41**, 1704 (2016).

[28] X. Zhang, S. Zhang, C. Chang, Y. Feng, Y. Li, N. Dong, K. Wang, L. Zhang, W. J. Blau, and J. Wang, Nanoscale **7**, 2978 (2015).

[29] M. Liu, X.-F. Jiang, Y.-R. Yan, X.-D. Wang, A.-P. Luo, W.-C. Xu, and Z.-C. Luo, Optics Communications https://doi.org/10.1016/j.optcom.2017.04.020, (2017).

[30] J. M. Soler, E. Artacho, J. D. Gale, A. Garca, J. Junquera, P. Ordejn, and D. Snchez-Portal, Journal of Physics: Condensed Matter **14**, 2745 (2002).

[31] N. Berkaine, E. Orhan, O. Masson, P. Thomas, and J. Junquera, Phys. Rev. B **83**, 245205 (2011).

[32] T. G. Pedersen, Phys. Rev. B **95**, 235419 (2017).

[33] J. P. Perdew, K. Burke, and M. Ernzerhof, Phys. Rev. Lett. **78**, 1396 (1997).

[34] A. N. Rudenko and M. I. Katsnelson, Phys. Rev. B **89**, 201408 (2014).

[35] Y. Jing, X. Zhang, and Z. Zhou, Wiley Interdisciplinary Reviews: Computational Molecular Science,**5** (2016).

[36] X. Peng, Q. Wei, and A. Copple, Phys. Rev. B **90**, 085402 (2014).

[37] B. Sa, Y.-L. Li, J. Qi, R. Ahuja, and Z. Sun, The Journal of Physical Chemistry C **118**, 26560 (2014).

[38] H. Liu, A. Neal, Z. Zhu, Z. Luo, X. Xu, D. Tomanek, and P. Ye, ACS Nano **8**, 4033 (2014).

[39] L. Liang, J. Wang, W. Lin, B. G. Sumpter, V. Meunier, and M. Pan, Nano Letters **14**, 6400 (2014).

[40] M. Ezawa, New Journal of Physics **16**, 115004 (2014).

[41] S. Shekarforoush, D. Shiri, and F. Khoeini, In preparation (2017).

[42] Y. Takimoto, F. D. Vila, and J. J. Rehr, The Journal of Chemical Physics **127**, 154114 (2007).

[43] R. Boyd, Nonlinear optics, 3rd Edition, Academic Press (2008).

[44] R. Chen, X. Zheng, and T. Jiang, Opt. Express **25**, 7507 (2017).

[45] D. Li, H. Jussila, L. Karvonen, G. Ye, H. Lipsanen, X. Chen, and Z. Sun, Scientific Reports **5**, 15899 (2015).

[46] H.-Y. Wu, Y. Yen, and C.-H. Liu, Applied Physics Letters **109**, 261902 (2016).

[47] X. Wang and S. Lan, Adv. Opt. Photon. **8**, 618 (2016).

[48] K.Wang, Y. Feng, C. Chang, J. Zhan, C.Wang, Q. Zhao, J. N. Coleman, L. Zhang, W. J. Blau, and J. Wang, Nanoscale **6**, 10530 (2014).